\documentclass[11pt]{article}
\usepackage[margin=0.95in]{geometry}
\usepackage{epsfig,ulem,amssymb,soul}

\usepackage{graphicx} 
\usepackage{epstopdf}
\usepackage{amsfonts,amsmath}
\usepackage{bm}
\usepackage{setspace}
\usepackage{comment}
\usepackage{siunitx}
\usepackage{mathrsfs}
\usepackage{multirow}
\usepackage{subcaption}
\usepackage[hang,flushmargin,bottom]{footmisc}
\usepackage{mathtools}
\usepackage{lipsum}
\usepackage{array}
\usepackage{enumerate}
\usepackage{ifthen}
\usepackage[affil-it]{authblk}
\usepackage{cite}
\usepackage{leftidx}
\usepackage{relsize}
\usepackage{braket}
\usepackage{float}

\usepackage[autostyle]{csquotes}
\usepackage[dvipsnames]{xcolor}
\newcommand{\etal}{\textit{et al}.~}

\providecommand{\keywords}[1]{\textbf{\textit{Keywords---}} #1}

\title{Multiple scattering simulation via physics-informed neural networks}
\date{}

\author[1]{Siddharth Nair}
\author[2]{Timothy F. Walsh}
\author[2]{Greg Pickrell}
\author{Fabio Semperlotti\thanks{To whom correspondence should be addressed. Email: fsemperl@purdue.edu }}
\affil[1]{Ray W. Herrick Laboratories, School of Mechanical Engineering,  Purdue University, West Lafayette, IN 47907, USA}
\affil[2]{Sandia National Laboratory, Albuquerque, NM 87185, USA}
\setstcolor{red}

\usepackage{lineno}

\begin{document}
\maketitle


\begin{abstract}

This work presents a physics-driven machine learning framework for the simulation of acoustic scattering problems. The proposed framework relies on a physics-informed neural network (PINN) architecture that leverages prior knowledge based on the physics of the scattering problem as well as a tailored network structure that embodies the concept of the superposition principle of linear wave interaction.
The framework can also simulate the scattered field due to rigid scatterers having arbitrary shape as well as high-frequency problems. Unlike conventional data-driven neural networks, the PINN is trained by directly enforcing the governing equations describing the underlying physics, hence without relying on any labeled training dataset. Remarkably, the network model has significantly lower discretization dependence and offers simulation capabilities akin to parallel computation. This feature is particularly beneficial to address computational challenges typically associated with conventional mesh-dependent simulation methods. The performance of the network is investigated via a comprehensive numerical study that explores different application scenarios based on acoustic scattering.

\noindent\keywords{Physics-informed neural network (PINN), Multiple scattering, Superposition principle, Wave propagation}
\end{abstract}

\section{Introduction}
\label{ssec: Introduction}

The fundamental concept of scattering holds a critical role in the general area of wave physics including, but not limited to, photonics \cite{bayati2020inverse, pestourie2018inverse}, geophysics \cite{speziale2014brillouin}, bio-medicine \cite{yaman2013survey}, acoustics \cite{wu2022physics, nair2023grids}, and particle physics \cite{faddeev1993quantum, amsler2008review}.
Within these fields, scattering finds applications in the most diverse areas including, but not limited to, non-destructive testing \cite{altpeter2002robust}, metrology \cite{krivosheev2022state}, advanced medical imaging \cite{bertero2006inverse, freudiger2008label}, remote sensing \cite{entekhabi1994solving}, material characterization \cite{waseda1984novel, carvalho2018application}, and inverse material design \cite{nair2023grids, kim2018deep}.    

Over the years, various numerical approaches have been developed to tackle the complex problem of scattering simulations. Finite element methods (FEM) \cite{volakis1994review, ihlenburg1998finite}, finite difference methods (FDM) \cite{morgan2013finite}, boundary element methods \cite{fahy2007sound}, and discontinuous Galerkin methods \cite{giorgiani2013high} have been among the most popular. Out of all these existing techniques, FEM distinguished itself for its ability to handle complex geometries and irregular boundaries. However, being a mesh-based technique, FEM suffers from a strong dependence on the discretization parameters. This dependence renders the FE models very computationally intensive when applied to scattering problems involving a large number of degrees of freedom, such as those for high frequency, multiple scattering, and multiscale applications \cite{thompson2006review, liu2003mesh}.
In particular, when dealing with multiple scattering scenarios, a notable challenge arises as the scale and number of scatterers increases. In these situations, FEM simulations become very computationally intensive or even intractable (i.e. exceeding the resources of available supercomputers). In fact, as the number of scattering elements increases the mesh requirements become more stringent (in order to capture the interactions between different scatterers) and the number of degrees of freedom increases very rapidly. Further studies have developed meshfree approaches like meshless local Petrov-Galerkin \cite{atluri1998new}, smoothed particle hydrodynamics \cite{monaghan1992smoothed}, kernel-based meshless method \cite{schaback2007convergence}, and element-free Galerkin method \cite{belytschko1994element} to address these limitations of FEM.
The ability of these methods to eliminate the need for complex mesh generation, which can be very time-consuming (particularly for problems involving irregular geometries or evolving domains) offers a significant advantage over mesh-based approaches. However, meshless methods often increase the computational costs, compared to mesh-based methods, due to the need for large numbers of nodes to accurately represent the domain \cite{nguyen2008meshless}. Moreover, ensuring stability and accuracy can be more complex in meshless approaches due to the lack of structured elements and their dependence on neighbor node connectivity, especially in problems involving discontinuities \cite{nguyen2008meshless, liew2011review}.

More recently, the use of machine learning as a tool to perform computations has seen a rapid increase, ultimately leading to the formulation of several models and techniques for a variety of applications. In the field of scattering simulations, examples include supervised learning-based deep neural networks (DNNs) to estimate optical scattering produced by nanoparticles \cite{peurifoy2018nanophotonic}, and the solution of the Poisson's equation in 2D domains with simple scatterer shapes \cite{shan2020study}. Convolutional neural networks (CNNs) and recurrent neural networks (RNNs) have been employed to solve the forward finite difference time domain problems \cite{yao2018machine}. Recent studies have also introduced semi-supervised \cite{nair2023grids} and unsupervised \cite{liu2018generative} deep learning models for material design applications in acoustics and optics. A major drawback of these conventional DNNs is the need for comprehensive labeled datasets to achieve effective network training. Furthermore, these trained DNNs are inherently incapable of guaranteeing physically consistent results. Hence, various engineering applications have shifted their focus towards physics-driven deep learning models that weakens the data dependence. This latter process has led to the development of the concept of physics-driven machine learning \cite{karpatne2022knowledge}, which aims at integrating the governing physics into conventional DNNs. 

In physics-driven machine learning, the integration of the physics of the problem into a learning algorithm amounts to introducing appropriate biases in the DNN that can guide the network prediction towards physically consistent solutions. Among the different approaches that can be leveraged to embed the physics of the problem within the network \cite{karniadakis2021physics}, one of the most effective and extensively studied concept is that of physics-informed neural network (PINN) \cite{raissi2019physics}. Since the dynamic behavior of most of the applications in the field of engineering is often described using governing laws expressed as a system of partial differential equations (PDEs), the ability of PINN to approximate numerical solutions of PDEs finds application in a wide range of domains. To-date, the literature shows that PINNs can address problems in a variety of fields, including fluid mechanics \cite{raissi2020hidden, jin2021nsfnets}, heat transfer \cite{zobeiry2021physics}, and solid mechanics \cite{samaniego2020energy} and more. Moreover, the use of PINNs offers two significant advantages over conventional DNN. First, PINNs can train without any need for labeled datasets, thus addressing the challenge of generating (either computationally or experimentally) extensive training data. Second, PINNs can learn to approximate results that are consistent with the underlying physics, therefore overcoming the tendency of conventional DNNs (trained on labeled datasets) to produce physically inconsistent predictions. In addition, and not less important, unlike the conventional simulation approaches (e.g. FEM, FDM), PINNs reduce discretization dependence. 

Recent studies have introduced PINNs for scattering applications across different fields. Chen \etal \cite{chen2020physics, chen2022physics} explored the problem of inverse parameter identification from optical scattered fields using PINNs. This work integrated the Maxwell equations into a PINN framework in order to extract the permittivity of the homogenized material based on the known scattered field. These parameters are crucial to characterize the scatterers in optical applications.

PINNs were also developed as solvers for transient forward problems involving the seismic wave equation \cite{karimpouli2020physics} and the full-wave inversion \cite{rasht2022physics} for geophysical applications. Song \etal \cite{song2021solving} developed a PINN to solve the frequency-domain acoustic wave equation for transversely isotropic (TI) media with a vertical axis of symmetry (VTI) for geophysics applications. Wang \etal \cite{wang2023acoustic} proposed a PINN model targeted to steady-state acoustic applications. The studies mentioned above developed PINN models to predict the scattered fields due to material heterogeneities whose influence was captured by an explicit material parameter within the governing PDE. 
In other terms, these previous studies investigated the scattered field due to scattering sources uniformly distributed across the domain, and typically employed homogenization techniques. However, a more complex and unexplored application of PINNs involves simulating the steady-state scattering response due to fully-resolved scatterers embedded within the domain. In this case, the scattered field is produced due to the wave interaction with the embedded scatterers, hence making it a function of the scatterer shape. 
It is important to note that, unlike the material parameters within the PDE, there is no equivalent parameter within the governing PDE that can explicitly account for the effect of the scatterer shape.
In this latter scenario, the presence of rigid scatterers within the unbounded domain introduces additional (both internal and external) boundary conditions. These boundary conditions are independent of the governing PDE, and they play a crucial role in shaping the behavior of the scattered field.

Therefore, the objective of this study is to establish a PINN framework capable of solving the scattered field generated by rigid scatterers embedded within an unbounded homogeneous domain.

\subsection{Major contributions}

The primary goal of this study is to develop a class of PINNs tailored to address a wide range of acoustic scattering problems involving arbitrary shaped scatterers, and high-frequency excitations. In this context, the term \enquote{arbitrary shape} indicates a wide range of scatterers with both regular and irregular shapes, while the term \enquote{high frequency} indicates wavelengths smaller than the characteristic size of the scatterer. More specifically, we focus on the development of PINNs to find approximate solutions to steady-state harmonic acoustic problems (governed by the Helmholtz equation) involving arbitrary-shaped rigid scatterers distributed in an unbounded (infinite) acoustic domain.
The key contributions of this study are strictly connected to the two essential components of the proposed PINN framework
\begin{enumerate}
    \item \textbf{Baseline-PINN ($b$-PINN):} A PINN architecture conceived to simulate acoustic scattering problems involving either single or (a small number of) multiple rigid scatterers of arbitrary shapes. In contrast to some existing models, such as those in the DeepXDE library \cite{lu2021deepxde} that are suitable for enforcing one basic internal boundary shape, the baseline-PINN is specifically designed to simulate fields generated by arbitrary scatterer shapes (introducing boundaries internal to the domain). Furthermore, the $b$-PINN architecture integrated with a tailored optimization procedure (a combination of gradient-based and second-order optimization) is well suited to simulate steady-state scattered fields at high frequencies (i.e. wavelength much shorter than the characteristic scatterer size). It is important to note that the $b$-PINN is also able to solve multi-scatterer configurations (although with some limitations on the total number of scattering elements). Contrarily to classical solution methods (e.g. FEM), in this latter case the performance of $b$-PINN is not limited by the number of degrees of freedom. Nevertheless, a limitation is imposed by the error propagation within the network \cite{jagtap2020conservative}, as discussed below. 

    \item \textbf{Superposition-PINN ($s$-PINN)}: A PINN architecture specifically conceived to address the limitations of $b$-PINNs when applied to large number of scatterers. Broadly speaking, the superposition-PINN is based on an architecture that leverages multiple $b$-PINNs. The uniqueness of this network lies in its dual-level integration of the underlying physics: 1) the $s$-PINN architecture embodies the superposition principle of linear acoustics, and 2) the governing PDE and boundary conditions are directly enforced through the network loss function. This approach addresses the limitation of $b$-PINNs by providing extensive control over the local distribution of scatterers within the multiple scatterer domain. Specifically, the $s$-PINN implements a domain partitioning approach such that individual $b$-PINNs can be used to simultaneously handle separate sub-domains.
    In addition, the individual $b$-PINNs within the overall $s$-PINN architecture naturally enable parallel processing by assigning each $b$-PINN training to separate computational cores. This parallelization approach can significantly reduce the total simulation time for multiple scattering studies as the $s$-PINN runtime is only dominated by the largest $b$-PINN and is not the sum of the runtimes of all the $b$-PINNs. 
\end{enumerate}

It follows that, by virtue of its reduced spatial discretization dependence and its parallelization capabilities, the resulting physics-driven machine learning-based forward solver can effectively reduce the computational cost associated with multiple scattering simulations compared to traditional methods (based on the solution of systems of differential equations). While PINNs exhibit lower discretization dependence during the training process, the ability of the trained PINNs to approximate solutions on a new discretization makes the prediction process discretization-independent. However, it is important to note that the primary focus of this work is on introducing the above-mentioned PINN methodologies to enable certain classes of simulations (such as multiple scattering), rather than optimizing the computational performance.

The paper is organized as follows. Section \ref{ssec: PINN_preliminaries} 
introduces some background on PINNs.
Section \ref{ssec: Single_scatterer} presents the general problem setup for single scatterers, elaborates on the implementation of the $b$-PINN, and reports and discusses the results of the trained $b$-PINNs. Thereafter, section \ref{ssec: Multiple_scatterer} presents the general problem setup for multiple scattering, elaborates on the implementation of the $s$-PINN, and reports and discusses the results of the trained $s$-PINNs.


\section{Basics of physics-informed neural network (PINN)}
\label{ssec: PINN_preliminaries}

Multiple approaches can be followed to construct a deep learning model capable of simulating the physical response of a system. Data-driven deep learning models can capture the physical response of the system by optimizing the neural network hyperparameters that lead to the minimization of the error between the training data and the network predictions. A major drawback of the data-driven approach is that it requires a large volume of training data and the predictions are often physically inconsistent (as the model can only capture the physical response represented by the training data). On the other hand, physics-informed neural networks (PINNs) belong to a class of deep learning algorithms that can integrate prior physical information about the problem to improve the performance of the learning algorithms. More specifically, PINNs can enforce the deep learning algorithm to learn the mathematical framework of the problem in the form of governing partial differential equations (PDEs) and constitutive equations. The two major advantages of using PINNs in comparison to conventional DNNs are: 1) PINNs present the capability to train without labeled datasets, and 2) PINNs can learn to predict physically consistent results.

In a general PINN framework, the unknown solution $\phi$ is computationally predicted by a neural network parameterized via a set of parameters $\theta=[\textbf{W}, \textbf{b}]$, where \textbf{W} and \textbf{b} are the sets of weights and biases of the network, respectively. The PINN approximation of the solution is $\hat{\phi}$ such that $\hat{\phi}(\theta) \approx \phi$.
Moreover, the PINN learns to predict $\hat{\phi}$ by finding the optimal $\theta$ by minimizing a loss function $\mathcal{L}(\theta)$ 
\begin{equation}
     \label{eqn: PINN_theta}
     \theta^* = \arg \min_{\theta} \mathcal{L}(\theta) 
\end{equation}
The PINN loss function can be represented as follows
\begin{equation}
     \label{eqn: PINN_loss}
     \mathcal{L}(\theta) = w_{\mathcal{B}}\mathcal{L}_{\mathcal{B}}(\theta) + w_{\mathcal{N}}\mathcal{L}_{\mathcal{N}}(\theta)
\end{equation}
where $w_{\mathcal{B}}$ and $w_{\mathcal{N}}$ are the weighting factors and
\begin{equation}
\begin{split}
     \label{eqn: PINN_LossPDEBC}
     \mathcal{L}_{\mathcal{B}}(\theta) &= \frac{1}{N_b} \sum^{N_b}_{i=1} \Big|\mathcal{B} (\hat{\phi},  \textbf{x}^i_{b}; \theta) - g^i \Big|^2 \\
     \mathcal{L}_{\mathcal{N}}(\theta) &=  \frac{1}{N_n} \sum^{N_n}_{i=1} \Big|\mathcal{N} (\hat{\phi},  \textbf{x}^i_{n}; \theta) \Big|^2
\end{split}
\end{equation}
where $\mathcal{B}$ represents the boundary condition, $\mathcal{N}$ represents the governing PDE, and $g$ represents the boundary data. Further, $\mathcal{L}_\mathcal{B}(\theta)$ and $\mathcal{L}_\mathcal{N}(\theta)$ are the mean square errors (MSE) of the residual of the boundary condition and the governing PDE, respectively. Here, $\textbf{x}_b$ is an arbitrary point sampled from a distribution of $N_b$ boundary training points and $\textbf{x}_n$ is an arbitrary point sampled from a distribution of $N_n$ domain collocation points.


\section{Acoustic scattering in the frequency domain}
\label{ssec: Single_scatterer}

This section introduces the benchmark problem used to test the $b$-PINN architecture. As previously indicated, this problem is based on the acoustic scattering due to rigid scatterers in the frequency domain. Initially, some theoretical preliminaries of acoustic scattering and the corresponding mathematical formulation are presented. Then, the concept of physics-informed neural network (PINN) formulated for the specific benchmark problem is introduced. Finally, the performance of the proposed network is studied through a series of numerical experiments.

\subsection{Problem description}
\label{ssec: problem_statement1}

\begin{figure}[h!]
	\centering
	\includegraphics[width=1.0\linewidth]{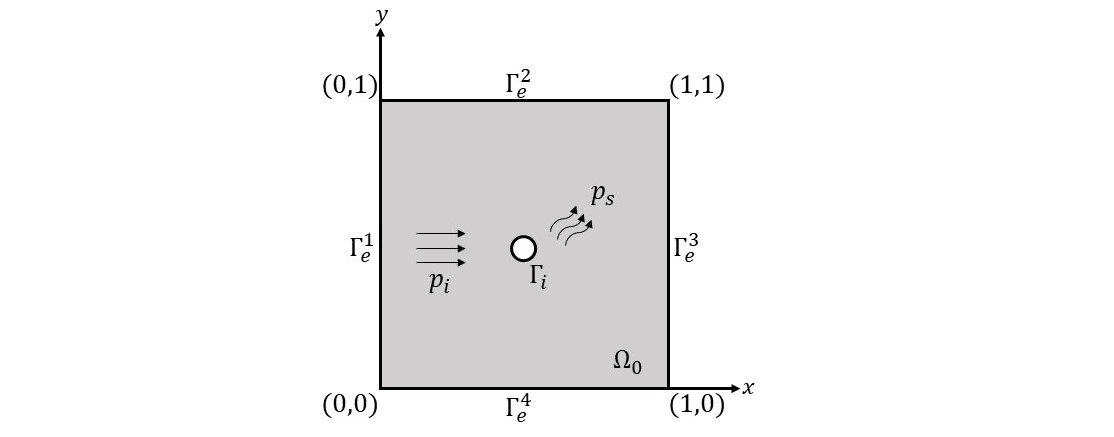}
	\caption{Schematic of a 2D square acoustic domain $\Omega_0$ of size $[0,1]m \times [0,1]m$. A rigid internal scatterer defines an internal boundary $\Gamma_i$, while the external boundaries of the domain are $\Gamma^1_e, \Gamma^2_e, \Gamma^3_e,$ and $\Gamma^4_e$. The incident pressure ($p_i$) and scattered pressure ($p_s$) are also indicated.}
	\label{fig: AcousticF_PINNSingle}
\end{figure}

Consider the classical problem of 2D acoustic scattering in air due to a single rigid scatterer. Assuming harmonic excitation and steady state conditions, the acoustic scattering problem is governed by the Helmholtz equation defined on a 2D domain $\Omega_0 \subset \mathbb{R}^2$  as shown in Fig.~\ref{fig: AcousticF_PINNSingle}
\begin{equation}
    \label{eqn: Helmholtz_equation}
    \mathcal{N}(p_s):=\nabla^2 p_s(\textbf{x}) + k^2 p_s(\textbf{x}) = 0, ~~~~~~~ \textbf{x} \in \Omega_0
\end{equation}
with the following external boundary conditions enforced on $\Gamma^1_e$, $\Gamma^2_e$, $\Gamma^3_e$, and $\Gamma^4_e$ 
\begin{equation}
\begin{split}
     \label{eqn: ext_BCs}
     \mathcal{B}^1_e (p_s)=g_1  ~~~ \textbf{x} \in \Gamma^1_e, ~~~~~~~ \mathcal{B}^2_e (p_s)=g_2  ~~~ \textbf{x} \in \Gamma^2_e \\
     \mathcal{B}^3_e (p_s)=g_3  ~~~ \textbf{x} \in \Gamma^3_e, ~~~~~~~ \mathcal{B}^4_e (p_s)=g_4  ~~~ \textbf{x} \in \Gamma^4_e 
\end{split}
\end{equation}
where $p_s$ is the scattered pressure, $\textbf{x}=(x,y)$ is the spatial coordinate, the wavenumber $\textbf{k} = \frac{2\pi f}{c_s} \hat{\textbf{e}}_k$ is a function of frequency $f$ and the speed of sound $c_s$, with unit normal vector $\hat{\textbf{e}}_k=[1,0]$, and boundary data $g_j$ for $j=1,2,3,4$. The differential operator $\mathcal{N}$ represents the governing PDE. The external boundary conditions prescribed by the boundary operators $\mathcal{B}^j_e$ for $j=1,2,3,4$ can take one of three possible forms: 1) Dirichlet boundary condition $\mathcal{B}^j_e=\textbf{I}$, 2) Neumann boundary condition $\mathcal{B}^j_e=\frac{\partial}{\partial \textbf{n}}$, or 3) impedance boundary condition $\mathcal{B}^j_e=\frac{\partial}{\partial \textbf{n}} + ik\textbf{I}$. 
To simulate an infinite acoustic domain, the Sommerfeld radiation condition (i.e. a throw-off boundary) is also enforced by means of the impedance boundary condition with $g_j=0$ (called the absorption boundary) on $\Gamma^1_e$, $\Gamma^2_e$, $\Gamma^3_e$, and $\Gamma^4_e$. 

While Eqs.~\ref{eqn: Helmholtz_equation}-\ref{eqn: ext_BCs} introduce the governing PDE and the external boundary conditions, the definition of the internal boundary condition (embedded within $\Omega_0$) of the rigid scatterer also plays a critical role. Moreover, the solution to Eq.~(\ref{eqn: Helmholtz_equation}) is determined by the pressure scattered by the rigid scatterer $\Gamma_i$ and indicated by $p_s= \mathfrak{Re}(p_s) + i\mathfrak{Im}(p_s)$.
For a unit amplitude ($p_0=1~Pa$) incident plane wave $p_i = p_0e^{-i\textbf{k} \cdot \textbf{x}}$ incident on a rigid scatterer $\Omega_0$, the boundary condition is enforced in the form of a Neumann  boundary (sound hard boundary) as follows
\begin{equation}
    \label{eqn: rigid_BC}
     \mathcal{B}_i (p_s):=\frac{\partial p_s(\textbf{x})}{\partial \textbf{n}}-ike^{-i\textbf{k} \cdot \textbf{x}}=0 ~~~~~~~ \textbf{x} \in \Gamma_{i}
\end{equation}
where $\mathcal{B}_i$ is the internal boundary operator and $\textbf{n}$ is the surface normal. Equations~\ref{eqn: Helmholtz_equation}-\ref{eqn: rigid_BC} provide the mathematical description of the forward acoustic scattering problem due to a single rigid scatterer $\Gamma_i$.

\begin{figure}[h!]
	\centering
	\includegraphics[width=1.0\linewidth]{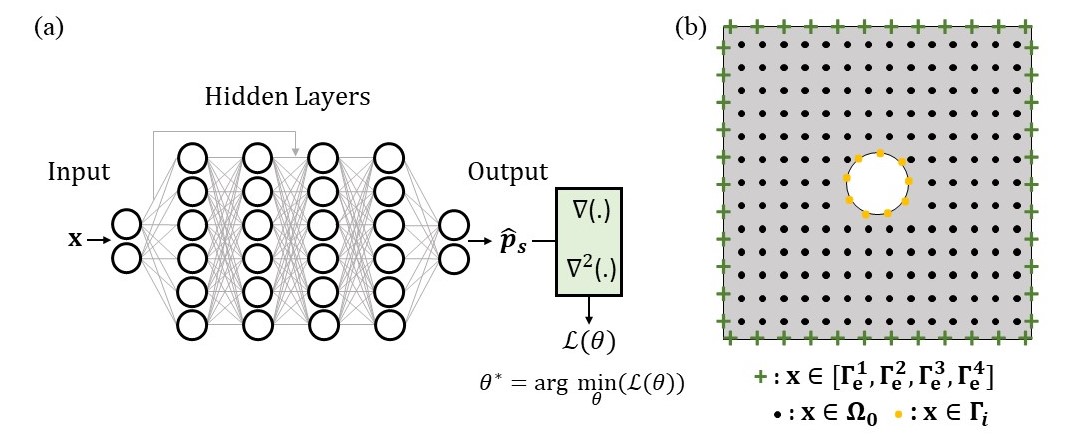}
	\caption{(a) High level schematic of the $b$-PINN architecture to simulate acoustic scattering. The spatial coordinates $\textbf{x}$ serve as the network input, and the network is trained to predict the scattered pressure $\hat{p}_s$ as output. (b) Schematic illustration of the coordinates used to train the $b$-PINN and sampled on the external boundaries ($\Gamma^1_e, \Gamma^2_e, \Gamma^3_e, \Gamma^4_e$), inside the acoustic domain ($\Omega$), and on the rigid internal boundary ($\Gamma_i$).}
	\label{fig: AcousticF_PINN}
\end{figure}

\subsection{Development of PINN for rigid body acoustic scattering}
\label{ssec: PINN_AcousticSingleSc}

Based on the elements discussed in sections \ref{ssec: PINN_preliminaries} and \ref{ssec: problem_statement1}, a $b$-PINN framework capable of solving the rigid body acoustic scattering problem is presented. Note that this section introduces the $b$-PINN parameters and the loss function, while a detailed discussion on the network architecture will be presented in the following section.

The $b$-PINN model consists of a neural network taking the spatial coordinates $\textbf{x}$ describing the acoustic domain as the network input (see Fig.~\ref{fig: AcousticF_PINN}(a)).
This deep learning model is designed to predict the unknown scattered pressure $p_s$ using the network parameterized by a set of parameters $\theta$. The approximate solution provided by the PINN $\hat{p}_s$ such that $\hat{p}_s \approx p_s$.
Moreover, the PINN learns to predict $\hat{p}_s$ by finding the optimal $\theta$ via minimization of the loss function $\mathcal{L}(\theta)$.
In the case of a single scatterer, the loss function $\mathcal{L}$ is defined to enforce Eqs.~\ref{eqn: Helmholtz_equation}-\ref{eqn: rigid_BC} as follows
\begin{equation}
    \label{eqn: loss_PINNtotal}
    \mathcal{L}(\mathbf{\theta}) =  w_{\mathcal{B}_i}\mathcal{L}_{\mathcal{B}_i}(\theta) + w_{\mathcal{B}_e}\mathcal{L}_{\mathcal{B}_e}(\theta) + w_{\mathcal{N}}\mathcal{L}_{\mathcal{N}}(\theta)
\end{equation}
where,
\begin{equation}
\begin{split}
     \label{eqn: loss_PINN_details}
     \mathcal{L}_{\mathcal{B}_i}(\theta) &=\frac{1}{N_{b_i}} \sum^{N_{b_i}}_{j=1}|\mathcal{B}_i(\hat{p}_s, \textbf{x}^j_{b_i};\theta)|^2 \\
     &= \frac{1}{N_{b_i}} \sum^{N_{b_i}}_{j=1} \Big|\frac{\partial  \hat{p}_s(\textbf{x}^j_{b_i})}{\partial \textbf{n}}-ike^{-i\textbf{k} \cdot \textbf{x}^j_{b_i}}  \Big|^2 \\
     \mathcal{L}_{\mathcal{B}_e}(\theta) &= \frac{1}{N_{b_e}} \sum^{N_{b_e}}_{j=1} \sum^4_{l=1} |\mathcal{B}_e(\hat{p}_s, \textbf{x}^{j,l}_{b_e};\theta)|^2 \\
     &= \frac{1}{N_{b_e}} \sum^{N_{b_e}}_{j=1} \sum^4_{l=1} \Big|\frac{\partial \hat{p}_s(\textbf{x}^{j,l}_{b_e})}{\partial \textbf{n}} + ik\hat{p}_s(\textbf{x}^{j,l}_{b_e})\Big|^2 \\
     \mathcal{L}_{\mathcal{N}}(\theta) &= \frac{1}{N_{n}} \sum^{N_n}_{j=1}|\mathcal{N}(\hat{p}_s, \textbf{x}^j_{n};\theta)|^2 \\
     &= \frac{1}{N_{n}} \sum^{N_n}_{j=1} \Big| \nabla^2 \hat{p}_s(\textbf{x}^j_{n}) + k^2 \hat{p}_s(\textbf{x}^j_{n}) \Big|^2
\end{split}
\end{equation}
where, $w_{\mathcal{B}_i}$, $w_{\mathcal{B}_e}$, and $w_{\mathcal{N}}$ are the weighting factors and  $\mathcal{L}_{\mathcal{B}_i}$, $\mathcal{L}_{\mathcal{B}_e}$, and $\mathcal{L}_{\mathcal{N}}$ are the mean square errors (MSE) of the residual of the rigid internal boundary conditions (Eq.~\ref{eqn: rigid_BC}), the external boundary conditions (Eq.~\ref{eqn: ext_BCs}), and the governing PDE (Eq.~\ref{eqn: Helmholtz_equation}), respectively. In addition, $\textbf{x}^j_{b_i} \in \Gamma_i$ is the $j^{th}$ coordinate from the $N_{b_i}$ internal boundary training points, $\textbf{x}^{j,l}_{b_e} \in \Gamma^l_e$ is the $j^{th}$ coordinate on the $l^{th}$ external boundary, where $l\in[1,2,3,4]$, from the $N_{b_e}$ external boundary training points, and $\textbf{x}^j_{n} \in \Omega_0$ is the $j^{th}$ coordinate from the $N_n$ number of domain collocation points. 

\subsection{Baseline-PINN architecture}
\label{ssec: baseline_PINN_arch}

With the loss functions fully defined to integrate the governing differential equation of the acoustic domain and the corresponding boundary conditions, we can address the setup of the network architecture. In the following, we first elaborate on $b$-PINN architecture, and later we cover the details for network training.    

While PINNs are commonly formed using fully connected neural network layers, recent studies \cite{qin2019data, karumuri2020simulator} have developed physics-driven deep learning models based on deep residual networks, also called ResNet architectures. These studies highlight the role of ResNets in improving training efficiency and prediction accuracy. Based on these previous findings, we choose a basic architecture for the PINN based on ResNet. Our ResNet architecture has $n_r$ residual blocks, where each residual block contains $n_l$ linear layers of $n_w$ neurons. Note that the residual block is an arrangement of linear neural network layers such that the input to the residual block is combined with its output deeper within the block for better information flow to the next stage of the DNN. Moreover, each linear layer is followed by an adaptive \textit{Sine} activation function. Adaptive activation functions include trainable parameters that are dynamically optimized during the training process to accelerate convergence and improve prediction accuracy in PINNs \cite{jagtap2020adaptive}.

A schematic illustration of the proposed architecture is shown in Fig.~\ref{fig: AcousticF_PINN}(a).
The architecture passes the input coordinate $\textbf{x}=(x,y)$ of size $2 \times 1$ through a linear layer of $n_w$ neurons, then the transformed input is passed through the residual blocks to generate $\hat{p}_s$ of size $2 \times 1$ as the output. Recall that, while $\hat{p}_s$ is a complex number with real and imaginary components, DNNs cannot operate with complex numbers. Therefore, the two components of $\hat{p}_s$ are stored as separate real numbers in the output vector, such that $\hat{p}_s=[\mathfrak{Re}(\hat{p}_s), \mathfrak{Im}(\hat{p}_s)]$.
In addition, as the solution approximation $\hat{p}_s$ contains both real and imaginary parts, the MSE losses in Eq.~\ref{eqn: loss_PINNtotal} are evaluated separately for the real and imaginary components. Automatic differentiation is used to calculate the gradients of $\hat{p}_s$ with respect to the input coordinate $\textbf{x}$ in the loss function.
The PINN architecture described above is called the \textit{baseline-PINN} or $b$-PINN.

\subsubsection{Training}
\label{ssec: training1}

In this section, we will discuss the network training details for the proposed $b$-PINN. It is important to note that while this section introduces the general parameters and their functions in network training, the specific details regarding the precise numerical values of these network parameters will be described as a part of the numerical experiments presented in the next section. 

In order to train the $b$-PINN, the spatial coordinates $\textbf{x}_{b_i}$, $\textbf{x}_{b_e}$, and $\textbf{x}_{n}$ serve as the input training points on $\Gamma_i$, the training points on $[\Gamma^1_e, \Gamma^2_e, \Gamma^3_e, \Gamma^4_e]$, and the collocation points in $\Omega_0$, respectively. Subsequently, the $b$-PINN is trained on mini-batches of the input data by sampling input coordinates $\textbf{x} \in [\textbf{x}_{b_i}, \textbf{x}_{b_e}, \textbf{x}_{n}]$ from their corresponding distribution of points as shown in Fig.~\ref{fig: AcousticF_PINN}(b). 
The network is trained using a two-step optimization process. Initially, we employ the Adam optimizer with a learning rate $l_r$ and train the network for $N_e$ number of epochs. Subsequently, we switch to the Limited-memory Broyden-Fletcher-Goldfarb-Shanno (L-BFGS) optimization method, a second-order optimization algorithm, to achieve faster convergence in the later stages of training. Moreover, the PINN is trained until $\mathcal{L}(\theta)$ converges by finding the optimal value of $\theta$ through backpropagation via automatic differentiation. 

The $b$-PINN is trained and implemented in Python 3.8 using Pytorch API on NVIDIA A100 Tensor Core GPU with 80GB memory.

\subsection{Numerical experiments and results}
\label{ssec: Results1}

This section studies the performance of the proposed $b$-PINN through a series of numerical experiments. More specifically, the evaluation focuses on assessing the capabilities of the proposed PINNs in simulating two applications. The first application focuses on examining the performance of the $b$-PINN in approximating the scattered acoustic field due to an arbitrary rigid scatterer. The second study assesses the ability of the $b$-PINN to perform high-frequency simulations, which is a common limitation of the existing PINNs.

The performance of the PINNs is measured by their ability to accurately approximate the scattered pressure fields. 
This assessment involves a direct comparison between the PINN predictions ($\hat{p}_s$) and their corresponding ground truth ($p_s$), which is obtained through finite element analysis.
In this process, the forward rigid body acoustic scattering problem is simulated using COMSOL Multiphysics$\textsuperscript{\textregistered}$, a FEM software, to evaluate $p_s$. To ensure the accuracy of the forward simulation model in FE, the rigid internal boundary is enforced through Neumann boundary conditions and the infinite acoustic boundaries are replicated by enforcing absorbing boundary conditions using a perfectly matched layer (PML). 
It is also important to note that since we are addressing a steady-state problem, the comparison between network predictions and the FEM ground truth is conducted at individual frequencies. To achieve this, for a characteristic length $a_0$ of a scatterer, forward simulations with a planar incident wave input are performed at different frequency (or wavelength) ranges corresponding to $ka_0<1$, $ka_0 \approx 1$, and $ka_0>1$, where the wavenumber $k = \frac{2\pi f}{c_s}$ is a function of frequency $f$ and speed of sound in air $c_s=343.21~m/s$. Broadly speaking, the frequencies are categorised into three ranges, namely, the low-frequency range ($ka_0<1$), the mid-frequency range ($ka_0 \geq 1$), the high-frequency range ($ka_0 >> 1$).  Although the current study deals with shapes of different sizes, we have evaluated a suitable reference of $a_0=0.1~m$ as an average characteristic length across the different scatterer shapes. Based on this, the current study categorizes $f<500~Hz$ ($ka_0 < 1$) as low-frequency range, $f=500~Hz-1~kHz$ ($ka_0 \geq 1$) as mid-frequency range, and $f>1~kHz$ ($ka_0 >> 1$) as high-frequency range. Given the relative straightforwardness of low-frequency applications in the context of a forward solver, this study emphasizes the use of the proposed PINNs for simulating rigid acoustic scattering specifically for applications in the mid- and high-frequency ranges.

Further, we introduce the following metrics to assess the overall performance of the networks
\begin{enumerate}
    \item \textbf{Relative $L_2$-error:} The following error metric is used to assess the relative prediction quality in an average sense
    \begin{equation}
        L_2 = \frac{\Big(\sum^N_{i=1} |p^i_s - \hat{p}^i_s|^2 \Big) ^{1/2}}{\Big(\sum^N_{i=1} |p^i_s|^2 \Big) ^{1/2}} 
    \end{equation}
    where, $p^i_s$ and $\hat{p}^i_s$ are the ground truth pressure value and the predicted pressure value, respectively, at the $i^{th}$ location in the corresponding fields. Additionally, $N$ represents the number of locations where the pressure fields are compared. The lower the $L_2$-error better the network prediction.

    \item \textbf{$R^2$-score:} The coefficient of determination also called the $R^2$-score is used to assess the accuracy of the network prediction. The $R^2$-score measures the average variation in the predicted field ($\hat{p}_s$) with respect to the ground truth field ($p_s$) as follows
    \begin{equation}
        R^2 = 1- \frac{\sum^N_{i=1} (p^i_s - \hat{p}^i_s)^2}{\sum^N_{i=1} (p^i_s - \bar{p}_s)^2} 
    \end{equation}
    The maximum score of $R^2_{max}=1$ is achieved when the prediction matches exactly with the ground truth. Hence, the closer the $R^2$-score is to $R^2_{max}$, the higher the prediction accuracy.
    
    \item \textbf{Point-wise error:} The following error metric assesses the point-wise prediction quality of the pressure field 
    \begin{equation}
        \mathcal{E}_{p} = |p_s - \hat{p}_s| 
    \end{equation}
    Here, $\mathcal{E}_{p}$ measures the absolute error to provide an error map highlighting variation between $p_s$ and $\hat{p}_s$ fields across the spatial domain. 
\end{enumerate}

\subsubsection{Application to arbitrary-shaped rigid scatterers}

In this first application, we study the performance of the $b$-PINN by investigating the prediction accuracy of the network for a single scatterer embedded within $\Omega_0$. More specifically, we choose to train the $b$-PINN to simulate the forward scattering problem for arbitrary rigid scatterer shapes. Table.~\ref{table:Archdetails_1} introduces the values of the network parameters chosen for this application denoted as Case 1.
\begin{table}[h!]
\setlength\tabcolsep{5pt}
	\begin{center}
		\begin{tabular}{ |c|c|c|c|c| } 
			\hline
			\multirow{1}*{\textbf{Parameter}} & \textbf{Case 1} & \textbf{Case 2} & \textbf{Case 2} & \textbf{Case 2}\\
            & & $f=2~kHz$& $f=3.5~kHz$& $f=5~kHz$\\
			\hline
			\hline  
			Hidden layer neurons ($n_w$) & 50 & 50 & 100 & 125 \\
            \hline   
			Number of residual blocks ($n_r$) & 2 & 2 & 2 & 3 \\
			\hline        
			Number of layers in each residual block ($n_l$) & 3 & 3 & 5 & 5 \\
			\hline
			Training points on $\Gamma_i$ ($N_{b_i}$) & 100 & 100 & 100 & 100 \\
            \hline
   		Training points on $\Gamma_e$ 
            ($N_{b_e}$) & 1000 & 1000 & 1000 & 1000 \\
            \hline
            Collocation points in $\Omega_0$ ($N_n$) & 10,000 & 10,000 & 10,000 & 10,000 \\
            \hline
            Learning rate ($l_r$) & 1e-4 & 1e-4 & 1e-4 & 1e-4 \\  
            \hline
		\end{tabular}
		\caption{Summary of key network parameters used to develop the $b$-PINN architecture for \textbf{Case 1:} arbitrary rigid scatterer simulation and \textbf{Case 2:} high-frequency simulation applications at $f=2~kHz, 3.5~kHz$, and $5~kHz$.}
		\label{table:Archdetails_1}
	\end{center}
\end{table}

\begin{figure}
	\centering
	\includegraphics[width=1.0\linewidth]{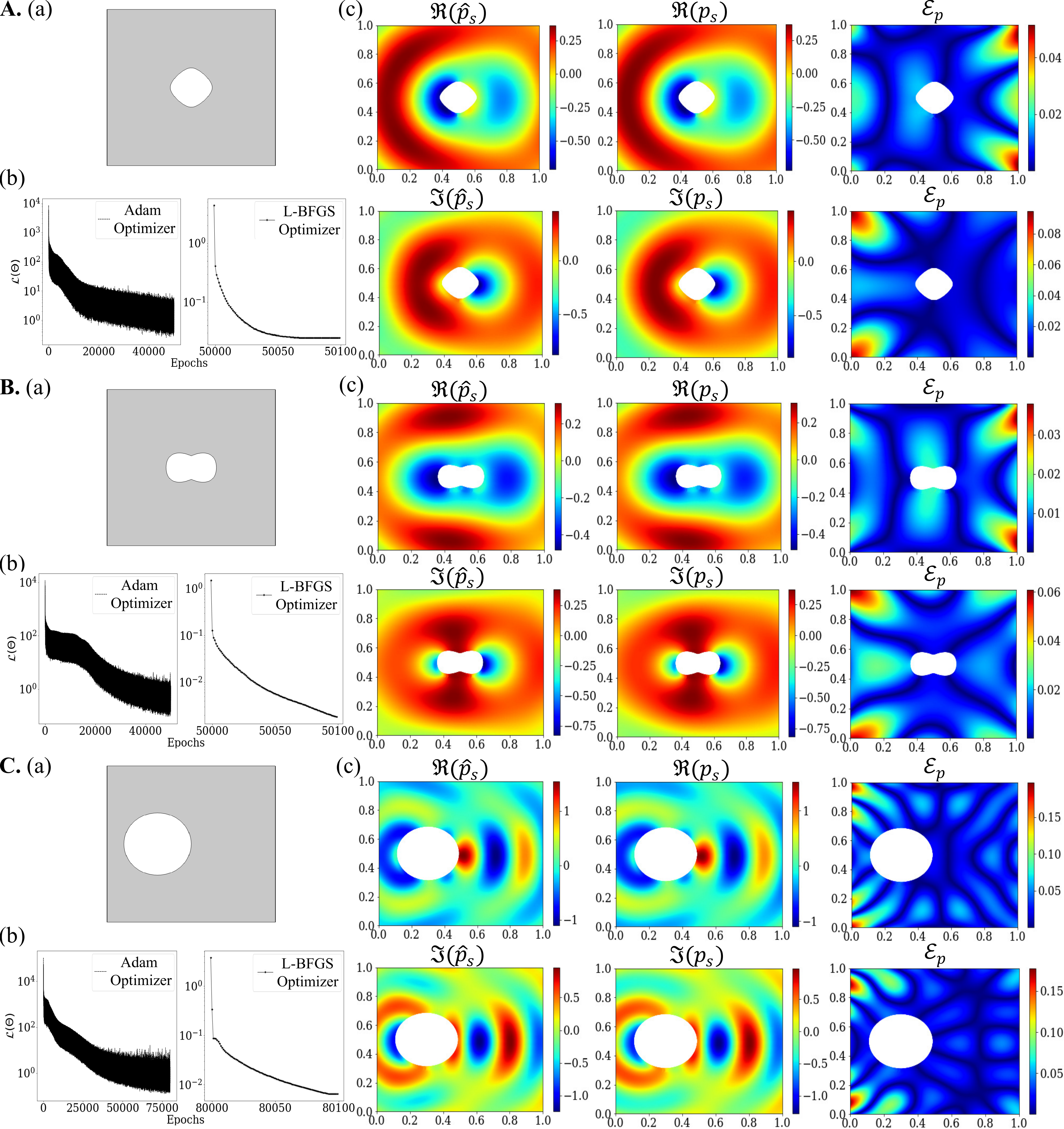}
	\caption{\textbf{Case 1- Arbitrary-shaped rigid scatterers:} Performance analysis of $b$-PINN for three scatterers described by an arbitrary shaped boundary (see (A)-(C)). The simulations in (A), (B), and (C) are performed for harmonic incident plane waves at $f=500~Hz$, $f=500~Hz$, and $f=1~kHz$, respectively. For each sample shape, the subplots (a), (b), and (c) represent: (a) the geometry of the acoustic domain and of the embedded rigid scatterer; (b) plots of the variation in the loss function with the training epochs (loss variations obtained using both Adam and L-BFGS optimizers are presented); (c) amplitude maps of the real (top) and imaginary (bottom) components of the predicted pressure field ($\hat{p}_s$), true pressure field ($p_s$), and the point-wise error ($\mathcal{E}$) between $\hat{p}_s$ and $p_s$.}
	\label{fig: Sc1_500Hz}
\end{figure}

Fig.~\ref{fig: Sc1_500Hz} illustrates the prediction accuracy of $b$-PINN for three random scatterer shapes. Within Fig.~\ref{fig: Sc1_500Hz}(A)-(C), the section marked (a) highlights the forward geometry model used for the simulation, (b) indicates the corresponding loss variation with training epochs using both Adam and L-BFGS optimizer, and (c) illustrates the prediction accuracy comparison for both real and imaginary scattered pressure components. Note that this is the general format used for all the figures in the rest of the numerical experiments as well. While Fig.~\ref{fig: Sc1_500Hz}(A) and (B) compare the prediction accuracy for scattered fields simulated by the $b$-PINN trained with $N_e=50,000$ epochs at $f=500~Hz$, Fig.~\ref{fig: Sc1_500Hz}(C) illustrates the performance of the $b$-PINN trained with $N_e=80,000$ epochs in simulating a scattered pressure field at $f=1~kHz$. The scattered fields generated for the shapes in Fig.~\ref{fig: Sc1_500Hz}(A) and (B) calculates $[L_2=0.0725$, $R^2=0.9937]$ and $[L_2=0.0749$, $R^2=0.9928]$, respectively. It is important to highlight that both the $L_2$-error and $R^2$-score are evaluated by averaging the $L_2$ and $R^2$ values of the real and imaginary scattered pressure components for each simulation. The high $R^2$-score and low $L_2$-error for the simulations in Fig.~\ref{fig: Sc1_500Hz}(A)and (B) highlights the ability of the $b$-PINN to accurately approximate the scattered field due to an arbitrary scatterer shape. Further, the performance evaluation for the simulation at $f=1~kHz$ in Fig.~\ref{fig: Sc1_500Hz}(C) calculates $[L_2=0.095$, $R^2=0.9909]$. This also showcases the ability of the $b$-PINN to accurately simulate the wavefield for different incident wave frequencies in the mid-frequency range. However, a comparison of the performance metrices between the simulations at $f=500~Hz$ and $f=1~kHz$ indicates a marginally higher prediction error at higher frequencies. This can be attributed to the inability of the same number of network weights ($n_w$) to capture finer wavefield features at higher frequencies.   

\subsubsection{Application to high-frequency scattering}

This application studies the performance of the $b$-PINN by investigating the prediction accuracy of the network for a single scatterer embedded within $\Omega_0$ at high frequency. More specifically, we choose to train the $b$-PINN to simulate the forward scattering problem for arbitrary rigid scatterer shapes at randomly chosen frequencies $f=2~kHz$ and $3.5~kHz$ in the high-frequency range. Table.~\ref{table:Archdetails_1} introduces the values of some of the essential network parameters chosen for this application denoted as Case 2.

\begin{figure}[h!]
	\centering
	\includegraphics[width=1.0\linewidth]{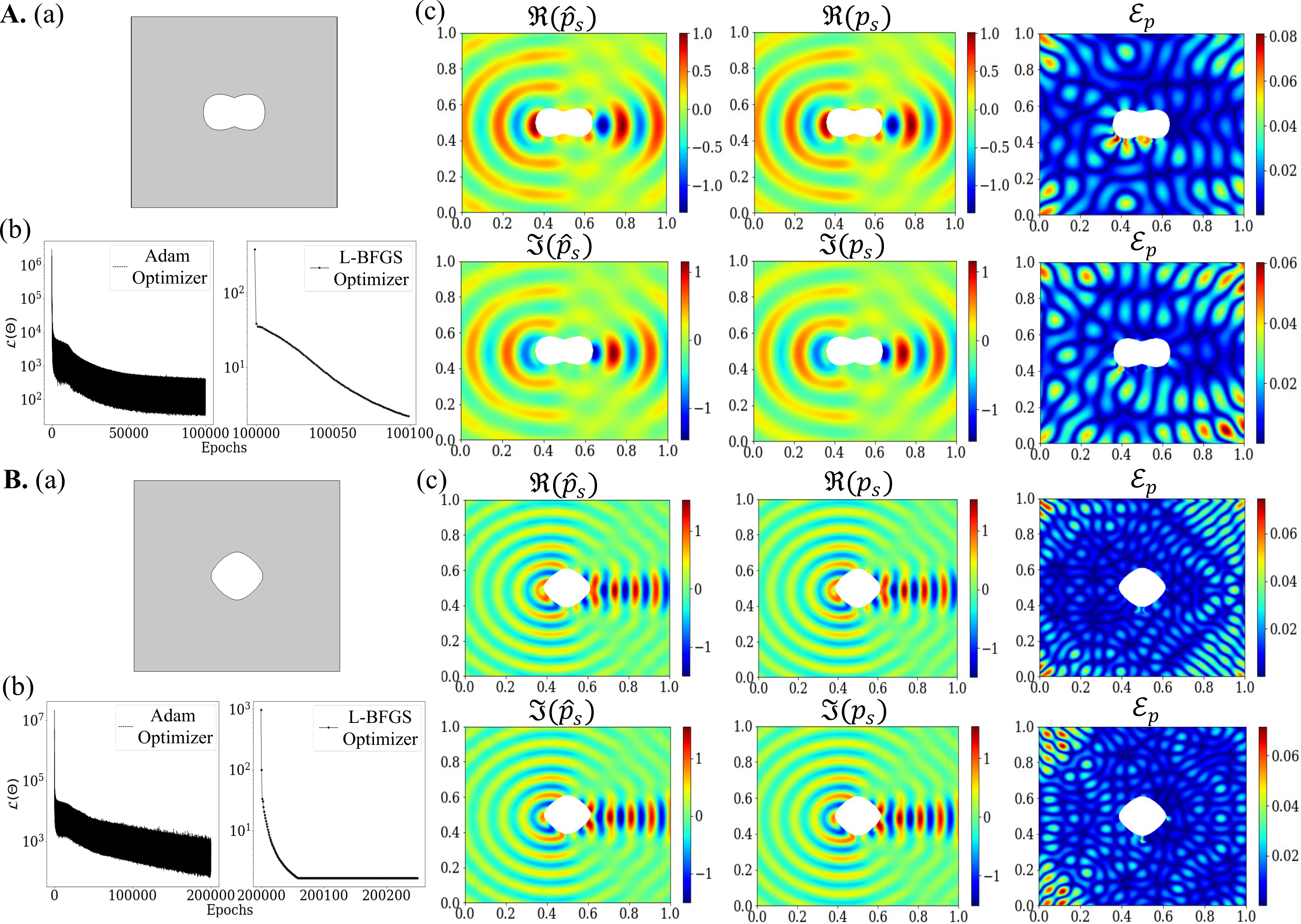}
	\caption{\textbf{Case 2- High frequency:} Performance analysis of $b$-PINN for high-frequency applications (see (A)-(B)). The simulations in (A) and (B) are performed for harmonic incident plane waves at $f=2~kHz$ and $f=3.5~kHz$, respectively. For each sample shape, the subplots (a), (b),
    and (c) represent: (a) the geometry of the acoustic domain and of the embedded rigid scatterer; (b) plots of the variation of the loss function with the training epochs (loss variations obtained using both Adam and L-BFGS optimizers are presented)); (c) amplitude maps of the real (top) and imaginary (bottom) components of the predicted pressure field ($\hat{p}_s$), true pressure field ($p_s$), and the point-wise error ($\mathcal{E}$) between $\hat{p}_s$ and $p_s$.}
	\label{fig: Sc1_HF}
\end{figure}

In the context of our discussion in the previous section, the increase in network prediction error with frequency is more evident in the high-frequency range. To address this, we observed that an increase in network prediction accuracy can be achieved by introducing more $n_w$ and adjusting $N_e$.
Therefore, the $b$-PINN is customized for specific frequencies by modulating $n_w$ and $N_e$, as detailed in Table.~\ref{table:Archdetails_1}.
Subsequently, the $b$-PINN is trained to simulate the scattered pressure field for an arbitrary shape at $f=2~kHz$ with $n_w=50$ and $N_e=100,000$ as shown in Fig.~\ref{fig: Sc1_HF}(A). Similarly, the $b$-PINN is also trained to simulate the scattered pressure field for another arbitrary shape at $f=3.5~kHz$ with $n_w=100$ and $N_e=200,000$ as shown in Fig.~\ref{fig: Sc1_HF}(B). While the performance metrices for Fig.~\ref{fig: Sc1_HF}(A) are calculated as $[L_2=0.0682, R^2=0.9953]$, the performance metrices for Fig.~\ref{fig: Sc1_HF}(B) are calculated as $[L_2=0.04675, R^2=0.9978]$. The low $L_2$-error and high $R^2$-score for both simulations highlight the key ability of the $b$-PINN to approximate scattered pressure fields in the high-frequency range. Moreover, the constant number of training and collocation points used for all the numerical experiments in the mid- and high-frequency ranges highlight the discretization-independence of the proposed PINN across frequency ranges. However, it is important to highlight that as we move to higher frequencies, the scattering simulations use up larger computational resources as we have to design deeper networks (Table.~\ref{table:Archdetails_1}) to capture extensive scattered field features at high frequencies. 

Furthermore, we conducted simulations at $f=5~kHz$ by training $b$-PINN for $N_e=200,000$ epochs by sampling from 10,000 collocation points and computed performance metrics to be $[L_2=0.2595, R^2=0.9327]$. As we study very high frequencies, i.e. $f \geq 5~kHz$ ($ka_0 \geq 10$), it is observed that an increased sampling distribution size is essential for better performance, thereby needing a finer sampling space to capture more detailed scattered field features. In summary, this study demonstrated the capability of the $b$-PINN to simulate in the high-frequency range. However, the prediction accuracy reduces as we transition to very high frequency values, as indicated by the high error metrics for the simulation at $f=5~kHz$.


\section{Multiple acoustic scattering}
\label{ssec: Multiple_scatterer}

This section introduces the benchmark problem used to develop and test the superposition-PINN or $s$-PINN architecture. First, some theoretical and mathematical preliminaries of multiple scattering in acoustics are presented. Then, the concept of physics-informed neural network (PINN) formulated for the specific benchmark problem is introduced. Finally, the performance of the proposed network is studied through a series of numerical experiments.

\subsection{Problem description}
\label{ssec: problem_statement2}

\begin{figure}[h!]
	\centering
	\includegraphics[width=1.0\linewidth]{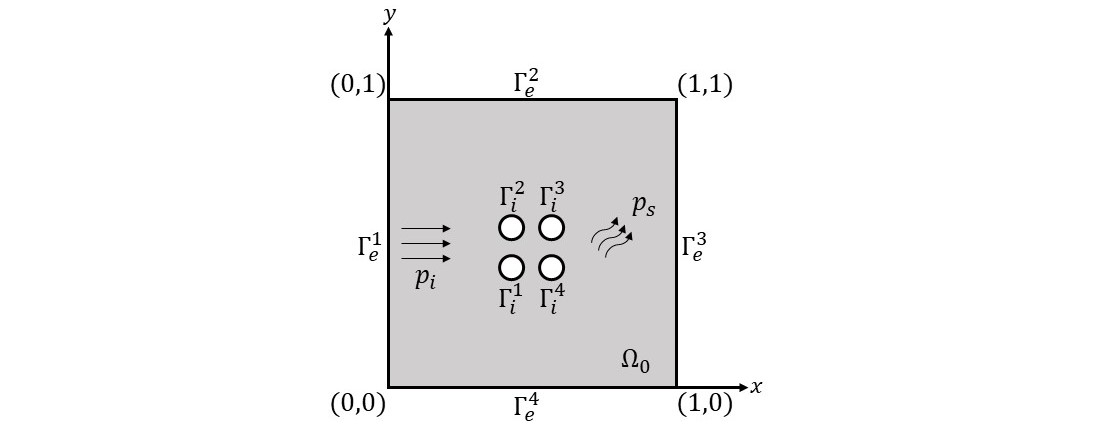}
	\caption{Schematic of 2D square acoustic domain $\Omega_0$ of size $[0,1]m \times [0,1]m$ embedded with multiple scatterers. The domain represents rigid internal boundary $\Gamma_i=\Gamma^1_i \cup \Gamma^2_i \cup \Gamma^3_i \cup \Gamma^4_i$ and external boundaries $\Gamma^1_e, \Gamma^2_e, \Gamma^3_e,$ and $\Gamma^4_e$. Further, the incident pressure ($p_i$) and scattered pressure ($p_s$) are also indicated.}
	\label{fig: AcousticF_PINNMultiple}
\end{figure}

A natural extension of the previous acoustic scattering problem involves the case of multiple rigid scatterers. Consider a domain $\Omega_0$ including multiple scatterers with rigid internal boundaries $\Gamma^1_i$, $\Gamma^1_i$, ...$\Gamma^{N_s}_i$. The scattered field radiated outward from each scatterer is $p^1_s, p^2_s, ...p^{N_s}_s$. A sample illustration including four scatterers ($N_s=4$) is shown in Fig.~\ref{fig: AcousticF_PINNMultiple}. According to the \textit{superposition principle}, the scattered pressure field at each point in the acoustic domain is the sum of the fields due to each scatterer \cite{koopmann1989method, pierce2019acoustics}. According to the theory of multiple scattering, the total scattered pressure ($p_s$) within $\Omega_0$ is
\begin{equation}
    \label{eqn: ps}
    p_s=\sum^{N_s}_{j=1} p^j_s
\end{equation}
where $p^j_s$ is the scattered field radiated outward from the $j^{th}$ scatterer.
By substituting Eq.~\ref{eqn: ps} in Eqs.~\ref{eqn: Helmholtz_equation}-\ref{eqn: rigid_BC}, we can generalize the governing equations to the case of $N_s$ scatterers. Leveraging the properties of linear differential operators, the governing Helmholtz equation for the multiple scattering case is
\begin{equation}
    \label{eqn: Helmholtz_equation_MS}
    \mathcal{N} \Big(\sum^{N_s}_{j=1} p^j_s \Big):= \nabla^2 \Big(\sum^{N_s}_{j=1} p^j_s(\textbf{x}) \Big) + k^2 \Big(\sum^{N_s}_{j=1} p^j_s(\textbf{x}) \Big) \equiv \sum^{N_s}_{j=1} \Big( \nabla^2  p^j_s(\textbf{x}) + k^2 p^j_s(\textbf{x}) \Big) = 0 
\end{equation}
The radiation boundary conditions are enforced on the external boundaries $\Gamma^1_e$, $\Gamma^2_e$, $\Gamma^3_e$, and $\Gamma^4_e$
\begin{equation}
\begin{split}
     \label{eqn: ext_BCs_MS}
     \mathcal{B}^1_e (\sum^{N_s}_{j=1} p^j_s):=& \sum^{N_s}_{j=1} \Big(\frac{\partial}{\partial \textbf{n}} p^j_s (\textbf{x}) + ik p^j_s(\textbf{x}) \Big) = 0  ~~~ \textbf{x}\in \Gamma^1_e\\
     \mathcal{B}^2_e (\sum^{N_s}_{j=1} p^j_s):=& \sum^{N_s}_{j=1} \Big(\frac{\partial}{\partial \textbf{n}} p^j_s (\textbf{x}) + ik p^j_s(\textbf{x}) \Big) = 0 ~~~ \textbf{x} \in \Gamma^2_e\\
     \mathcal{B}^3_e (\sum^{N_s}_{j=1} p^j_s):=& \sum^{N_s}_{j=1} \Big(\frac{\partial}{\partial \textbf{n}} p^j_s (\textbf{x}) + ik p^j_s(\textbf{x}) \Big) = 0 ~~~ \textbf{x} \in \Gamma^3_e\\
     \mathcal{B}^4_e (\sum^{N_s}_{j=1} p^j_s):=& \sum^{N_s}_{j=1} \Big(\frac{\partial}{\partial \textbf{n}} p^j_s (\textbf{x}) + ik p^j_s(\textbf{x}) \Big) = 0 ~~~ \textbf{x} \in \Gamma^4_e\\
\end{split}
\end{equation}
The Neumann boundary conditions are imposed on the rigid internal boundaries of each scatterer
\begin{equation}
\begin{split}
    \label{eqn: rigid_BC_MS}
     \mathcal{B}_i \Big( \sum^{N_s}_{j=1} p^j_s (\textbf{x}) \Big) :=& \frac{\partial}{\partial \textbf{n}} \Big (\sum^{N_s}_{j=1} p^j_s (\textbf{x}) \Big) -ike^{-i\textbf{k} \cdot \textbf{x}}=0 ~~~~~~~ \textbf{x} \in \Gamma_{i} \\
     :=& \sum^{N_s}_{j=1} \frac{\partial p^j_s (\textbf{x})}{\partial \textbf{n}} -ike^{-i\textbf{k} \cdot \textbf{x}}=0
\end{split}
\end{equation}
where $\Gamma_i$ represents the ensemble of rigid internal boundaries such that $\Gamma_i=\Gamma^1_i \cup \Gamma^2_i...\cup \Gamma^{N_s}_i$. In summary, Eqs.~\ref{eqn: Helmholtz_equation_MS}-\ref{eqn: rigid_BC_MS} provide the mathematical description of the forward multiple scattering problem.


\subsection{Development of PINN for multiple scattering}
\label{ssec: PINN_AcousticMultipleSc}

Based on the discussions in sections \ref{ssec: PINN_preliminaries} and \ref{ssec: problem_statement2}, this section presents a neural network framework, with a parameter set $\theta$, capable of solving the multiple scattering problem in acoustics.
For the multiple scattering case, the loss function $\mathcal{L}$ is defined to enforce Eqs.~\ref{eqn: Helmholtz_equation_MS}-\ref{eqn: rigid_BC_MS} as follows
\begin{equation}
    \label{eqn: loss_PINNtotal_MS}
    \mathcal{L}(\mathbf{\theta}) =  w'_{\mathcal{B}_i}\mathcal{L}'_{\mathcal{B}_i}(\theta) + w'_{\mathcal{B}_e}\mathcal{L}'_{\mathcal{B}_e}(\theta) + w'_{\mathcal{N}}\mathcal{L}'_{\mathcal{N}}(\theta)
\end{equation}
where, $\mathcal{L}'_{\mathcal{B}_i}$, $\mathcal{L}'_{\mathcal{B}_e}$, and $\mathcal{L}'_{\mathcal{N}}$ are the mean square errors (MSE) of the residual of the rigid internal boundary conditions (Eq.~\ref{eqn: rigid_BC_MS}), the external boundary conditions (Eq.~\ref{eqn: ext_BCs_MS}), and the governing PDE (Eq.~\ref{eqn: Helmholtz_equation_MS}), with $w'_{\mathcal{B}_i}$, $w'_{\mathcal{B}_e}$, and $w'_{\mathcal{N}}$ as the corresponding weighting factors. Here,
\begin{equation}
\begin{split}
     \label{eqn: loss_PINN_MS_details}
     \mathcal{L}'_{\mathcal{B}_i}(\theta) &=\frac{1}{N_{b_i}} \sum^{N_{b_i}}_{j=1} \Big| \mathcal{B}_i \Big(\sum^{N_s}_{q=1}\hat{p}^q_s, \textbf{x}^j_{b_i};\theta \Big) \Big|^2 \\
     &= \frac{1}{N_{b_i}} \sum^{N_{b_i}}_{j=1} \Big| \frac{\partial}{\partial \textbf{n}} \Big( \sum^{N_s}_{q=1} \hat{p}^q_s(\textbf{x}^j_{b_i}) \Big)-ike^{-i\textbf{k} \cdot \textbf{x}^j_{b_i}}  \Big|^2 \\
     \mathcal{L}'_{\mathcal{B}_e}(\theta) &= \frac{1}{N_{b_e}} \sum^{N_{b_e}}_{j=1} \sum^4_{l=1} \Big| \mathcal{B}_e \Big( \sum^{N_s}_{q=1}\hat{p}^q_s, \textbf{x}^{j,l}_{b_e};\theta \Big) \Big|^2 \\
     &= \frac{1}{N_{b_e}} \sum^{N_{b_e}}_{j=1} \sum^4_{l=1} \Big| \sum^{N_s}_{q=1} \Big( \frac{\partial \hat{p}^q_s(\textbf{x}^{j,l}_{b_e})}{\partial \textbf{n}} + ik\hat{p}^q_s(\textbf{x}^{j,l}_{b_e}) \Big) \Big|^2 \\
     \mathcal{L}'_{\mathcal{N}}(\theta) &= \frac{1}{N_{n}} \sum^{N_n}_{j=1} \Big| \mathcal{N} \Big( \sum^{N_s}_{q=1}\hat{p}^q_s, \textbf{x}^j_{n};\theta \Big) \Big|^2 \\
     &= \frac{1}{N_{n}} \sum^{N_n}_{j=1} \Big| \sum^{N_s}_{q=1} \Big( \nabla^2 \hat{p}^q_s(\textbf{x}^j_{n}) + k^2 \hat{p}^q_s(\textbf{x}^j_{n}) \Big) \Big|^2
\end{split}
\end{equation}


\subsection{Superposition-PINN architecture}
\label{ssec: superposition_PINN}

\begin{figure}[h!]
	\centering
	\includegraphics[width=1.0\linewidth]{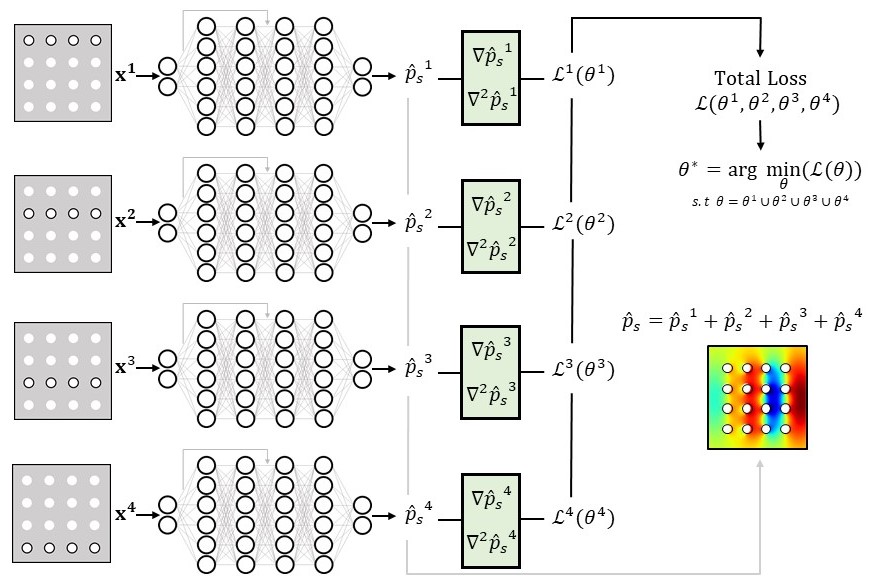}
	\caption{A schematic representation of the $s$-PINN architecture to simulate acoustic scattering in domains with multiple scatterers. In particular, this framework simulates a $4 \times 4$ lattice by simultaneously training four $b$-PINNs, each containing four rigid scatterers. Every $j^{th}$ $b$-PINN is trained to predict the scattered pressure $\hat{p}^j_s$ for the corresponding input coordinates $\textbf{x}^j$ by finding the optimal parameter set $\theta^j$ through minimization of the total loss $\mathcal{L}(\theta)$ which is a function of the loss components $\mathcal{L}^j(\theta^j)$ evaluated by the individual $b$-PINN. Finally, based on the concept of superposition in linear acoustics, the scattered pressure is evaluated as the sum of the scattered pressures simulated by the individual $b$-PINNs, i.e. $\hat{p}_s=\hat{p}^1_s+\hat{p}^2_s+\hat{p}^3_s+\hat{p}^4_s$.}
	\label{fig: SuperPINN_Arch}
\end{figure}

The $b$-PINN architecture (Fig.~\ref{fig: AcousticF_PINN}(a)), introduced in section \ref{ssec: Single_scatterer}, is capable of calculating the scattered pressure field due to either single or multiple scatterers by minimizing the loss functions presented in Eq.~\ref{eqn: loss_PINNtotal} (for the single scattering case) and in Eq.~\ref{eqn: loss_PINNtotal_MS} (for the multiple scattering case). In a multiple scattering scenario, our numerical results revealed that the prediction accuracy of the $b$-PINN decreases as the number of scatterers increases (see the comparative study in section \ref{ssec: Results2}). This reduction in accuracy, accompanied also by an increase in computational time, could be attributed to the limited ability of the single network $b$-PINN to minimize $\mathcal{L}$.
Indeed, when the number of scatterers increases, the high prediction error observed during the initial network training phase, as characterized by high $\mathcal{L}'_{{\mathcal{B}_i}}$ values in $\mathcal{L}$, propagates across the spatial domain. Consequently, there is a limited ability to control the error propagation while using a single network. To address this issue, we introduce a complementary PINN architecture, that leverages the superposition principle in linear acoustics; in the following, this network will be referred to as the \textit{superposition-PINN} or $s$-PINN.

The main goal of the $s$-PINN architecture is to overcome the limitations of the $b$-PINN when used in the context of multiple scattering. 
The basic concept at the foundation of $s$-PINN is to leverage a domain decomposition approach such that separate $b$-PINNs can be used simultaneously to obtain the response in each scatterer sub-domain. Subsequently, the scattered fields approximated by the individual $b$-PINNs are integrated using the superposition principle. Using this domain decomposition approach, each $b$-PINN within the $s$-PINN is constrained to simulate only a specific subset of the scatterers, thereby preventing high error propagation within each $b$-PINN. In a multiple scattering scenario, an increase in the total number of scatterers is accommodated by introducing additional $b$-PINNs without increasing the maximum number of scatterers each $b$-PINN can handle. Consequently, the $s$-PINN is not affected by the inaccuracies due to the error propagation. However, the scalability of $s$-PINN depends on the number of $b$-PINNs that can be simulated simultaneously on a given hardware configuration. The rest of this section elaborates on the implementation of the $s$-PINN.

%
We define a scatterer sub-domain as an acoustic domain $\Omega_0$ including $n_s$ scatterers, which represent a subset of the total number of scatterers $N'_s$. The number of scatterers in a scatterer sub-domain is typically $n_s \geq 1$, but it should be chosen so that the $b$-PINNs can provide predictions with the desired accuracy without deteriorating the computational cost to train the network. Fig.~\ref{fig: SuperPINN_Arch} illustrates a $s$-PINN architecture to simulate a lattice of $N'_s=16$ rigid scatterers divided across four sub-domains, each containing $n_s=4$ rigid scatterers. The $s$-PINN learns to approximate the total scattered field by simultaneously training all the $b$-PINNs. A $s$-PINN with $j$ $b$-PINNs is parameterized by a training parameter set $\theta$ such that $\theta=\theta^1 \cup \theta^2...\cup \theta^j$, where $\theta^j$ denotes the parameter set corresponding to the $j^{th}$ $b$-PINN. Further, the optimal $\theta$ is obtained by minimizing the superposition theorem-based total loss $\mathcal{L}$ defined in Eq.~\ref{eqn: loss_PINNtotal_MS}. Note that $N_s$ in Eq.~\ref{eqn: loss_PINNtotal_MS} is generalized as the number of $b$-PINNs within the $s$-PINN. Although $N_s$ was defined as the total number of scatterers in section \ref{ssec: problem_statement2}, in $s$-PINN $N_s$ can be less than or equal to the total number of scatterers, i.e. $N_s \leq N'_s$. After training, each $b$-PINN learns to predict the corresponding scattered field $\hat{p}^j_s$, where the index $j$ represents $j^{th}$ sub-domain. Thereafter, based on Eq.~\ref{eqn: ps}, the total scattered pressure ($\hat{p}_s$) due to the $N'_s$ scatterers is evaluated as the superposition of the scattered fields predicted by $N_s$ $b$-PINNs, i.e. $\hat{p}_s = \sum^{N_s}_{j=1} \hat{p}^j_s$.

To further clarify the basic principle of $s$-PINN, we present the example in Fig.~\ref{fig: SuperPINN_Arch}. Consider the $4 \times 4$ lattice of scatterers ($N'_s=16$) as described above. The $s$-PINN contains $N_s=4$ $b$-PINNs corresponding to an equal number of sub-domains, each containing $n_s=4$ rigid scatterers. In this approach, individual $b$-PINN is responsible for simulating the scattered field due to a subset of four scatterers within $\Omega_0$, as depicted in Fig.~\ref{fig: SuperPINN_Arch}. Alternatively, each $b$-PINN enforces the rigid internal boundary condition for the selected number of scatterers ($n_s=4$), as highlighted in bold within the scatterer sub-domains input to the individual $b$-PINNs in Fig.~\ref{fig: SuperPINN_Arch}. The proposed approach can effectively address the challenges of simulating multiple scattering by reducing the error propagation within each $b$-PINN. This is a significant improvement compared to training a single $b$-PINN to simulate the scattered field due to all 16 rigid scatterers. Finally, by applying the superposition theorem in linear acoustics, the trained $s$-PINN can estimate the total scattered pressure field $\hat{p}_s$ as the summation of the scattered fields $\hat{p}^1_s, \hat{p}^2_s, \hat{p}^3_s, \hat{p}^4_s$ approximated by the $b$-PINNs corresponding to the scatterer sub-domains with input coordinates $\textbf{x}^1, \textbf{x}^2, \textbf{x}^3, \textbf{x}^4$, i.e. $\hat{p}_s = \hat{p}^1_s + \hat{p}^2_s + \hat{p}^3_s + \hat{p}^4_s$.

\subsubsection{Training}
\label{ssec: training2}

\begin{figure}[h!]
	\centering
	\includegraphics[width=1.0\linewidth]{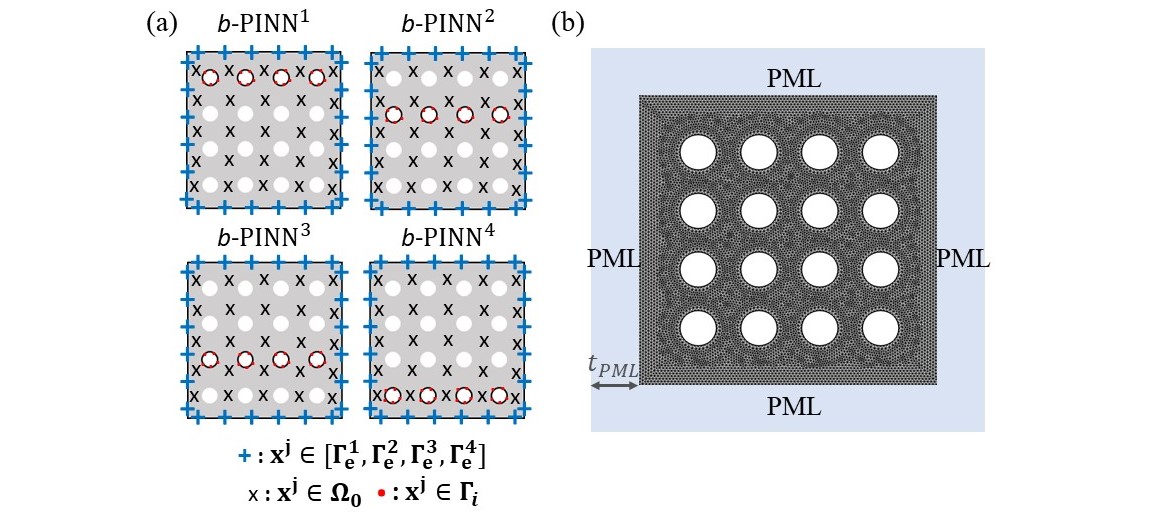}
	\caption{(a) Schematic illustration of the coordinates sampled on the external boundaries ($\Gamma^1_e, \Gamma^2_e, \Gamma^3_e, \Gamma^4_e$), inside the acoustic domain ($\Omega$), and rigid internal boundary ($\Gamma_i$) for training a $s$-PINN with 4 scatterer sub-domains. (b) The meshed geometry for a sample lattice with multiple rigid scatterers enforcing PML boundaries for finite element (FE) analysis.}
	\label{fig: Sampling_train}
\end{figure}

The $s$-PINN learns by simultaneously training $N_s$ $b$-PINNs. Every $j^{th}$ $b$-PINN is trained with the spatial coordinates $\textbf{x}^j_{b_i}$, $\textbf{x}_{b_e}$, and $\textbf{x}_{n}$ as the input training points on $\Gamma^j_i$, the training points on $[\Gamma^1_e, \Gamma^2_e, \Gamma^3_e, \Gamma^4_e]$, and the collocation points in $\Omega_0$, respectively. Subsequently, the $b$-PINN is trained on mini-batches of the input data by sampling input coordinates $\textbf{x}^j \in [\textbf{x}^j_{b_i}, \textbf{x}_{b_e}, \textbf{x}_{n}]$ from their corresponding distribution of points. For example, Fig.~\ref{fig: Sampling_train}(a) illustrates the sampling strategy for a $s$-PINN with $N_s=4$ $b$-PINNs. Note that within the inputs to every $j^{th}$ $b$-PINN, only the coordinates of the $\Gamma_i$ of the rigid scatterers vary, while the $\Gamma_e$ and $\Omega_0$ remain the same for all $b$-PINNs. The $s$-PINN is also trained with a learning rate $l_r$ using the Adam optimizer followed up with the L-BFGS optimization method. Moreover, the $s$-PINN is trained until $\mathcal{L}'_{{\mathcal{B}_i}}$, $\mathcal{L}'_{{\mathcal{B}_e}}$, and $\mathcal{L}'_{\mathcal{N}}$ converges by finding the optimal value of $\theta = \theta^1 \cup \theta^2...\theta^j$ for all the $b$-PINNs through backpropagation. 

The $s$-PINN is trained and implemented in Python 3.8 using Pytorch API on NVIDIA A100 Tensor Core GPU with 80GB memory.

\subsection{Numerical experiments and results}
\label{ssec: Results2}

This section explores and evaluates the performance of the proposed $s$-PINN architecture through a series of numerical experiments. The ability of the $s$-PINN to address multiple scattering scenarios is thoroughly examined, with a specific focus on its effectiveness in simulating the scattered acoustic pressure fields within acoustic lattice structures. Acoustic lattices (such as those at the basis of acoustic metamaterials) play an important role in practical applications because they are typically used to achieve wavefront manipulation and control. These structures often employ scatterers in periodic arrangements forming grids or lattices. The specific characteristics of the individual scatterers (e.g. shape, size, etc.) and of the grid (e.g. spacing, arrangement, etc.) are critical to determine the equivalent medium properties of the resulting assembly. These properties can be targeted and tailored to manipulate the acoustic wavefront as it propagates through the material. The periodic lattice example is selected to present the characteristics of the $s$-PINN.

As introduced in section \ref{ssec: Results1}, the performance assessment of PINNs involves a direct comparison between the PINN predictions ($\hat{p}_s$) and their corresponding ground truth ($p_s$), which is obtained through finite element analysis.
In this process, the forward rigid body acoustic scattering problem is simulated using COMSOL Multiphysics$\textsuperscript{\textregistered}$ to evaluate $p_s$. The FEM simulation setup for a sample multiple scattering scenario is shown in Fig.~\ref{fig: Sampling_train}(b). 

In these numerical experiments, we investigate the performance of the proposed networks when simulating acoustic lattices in the mid-frequency range. Specifically, we assess the performance of both the $b$-PINN and $s$-PINN for the same multiple scattering simulations. Additionally, we compare the performances of both networks to emphasize the significance of the superposition network.

\subsubsection{Multiple scattering simulation using $b$-PINN}

\begin{table}[ht]
    \centering
    \begin{tabular}{|c|c|c|c|c|c|}
        \hline 
        \multicolumn{1}{|c|}{\textbf{Parameter}}  & \multicolumn{1}{c|}{\textbf{Case 3 $b$-PINN}}  & \multicolumn{4}{c|}{\textbf{Case 3 $s$-PINN}} \\ \cline{3-6} 
         $4 \times 1$ lattice&   & PINN$^1$ & PINN$^2$ & PINN$^3$ & PINN$^4$ \\
        \hline
        \hline
		$n_w$ & 100 & 100 & 100 & 100 & 100 \\
        \hline   
		$n_r$ & 2 & 2 & 2 & 2 & 2 \\
		\hline        
		$n_l$ & 5 & 4 & 4 & 4 & 4 \\
		\hline
		$N_{b_i}$ & 100 & 100 & 100 & 100 & 100 \\
        \hline
   	$N_{b_e}$ & 1000 & 1000 & 1000 & 1000 & 
        1000 \\
        \hline
        $N_n$ & 10,000 & 10,000 & 10,000 & 10,000 & 10,000 \\
        \hline
        \multicolumn{1}{|c|}{$l_r$}  & \multicolumn{1}{c|}{1e-3, 1e-4}  & \multicolumn{4}{c|}{1e-3, 1e-4} \\
        \hline
    \end{tabular}
    \caption{Summary of key network parameters used to develop the $b$-PINN and $s$-PINN architecture for multiple scattering simulation due to a $4 \times 1$ scatterer lattice at $f=500~Hz$.}
    \label{table: Table_2}
\end{table}

\begin{figure}[h!]
	\centering
	\includegraphics[width=1.0\linewidth]{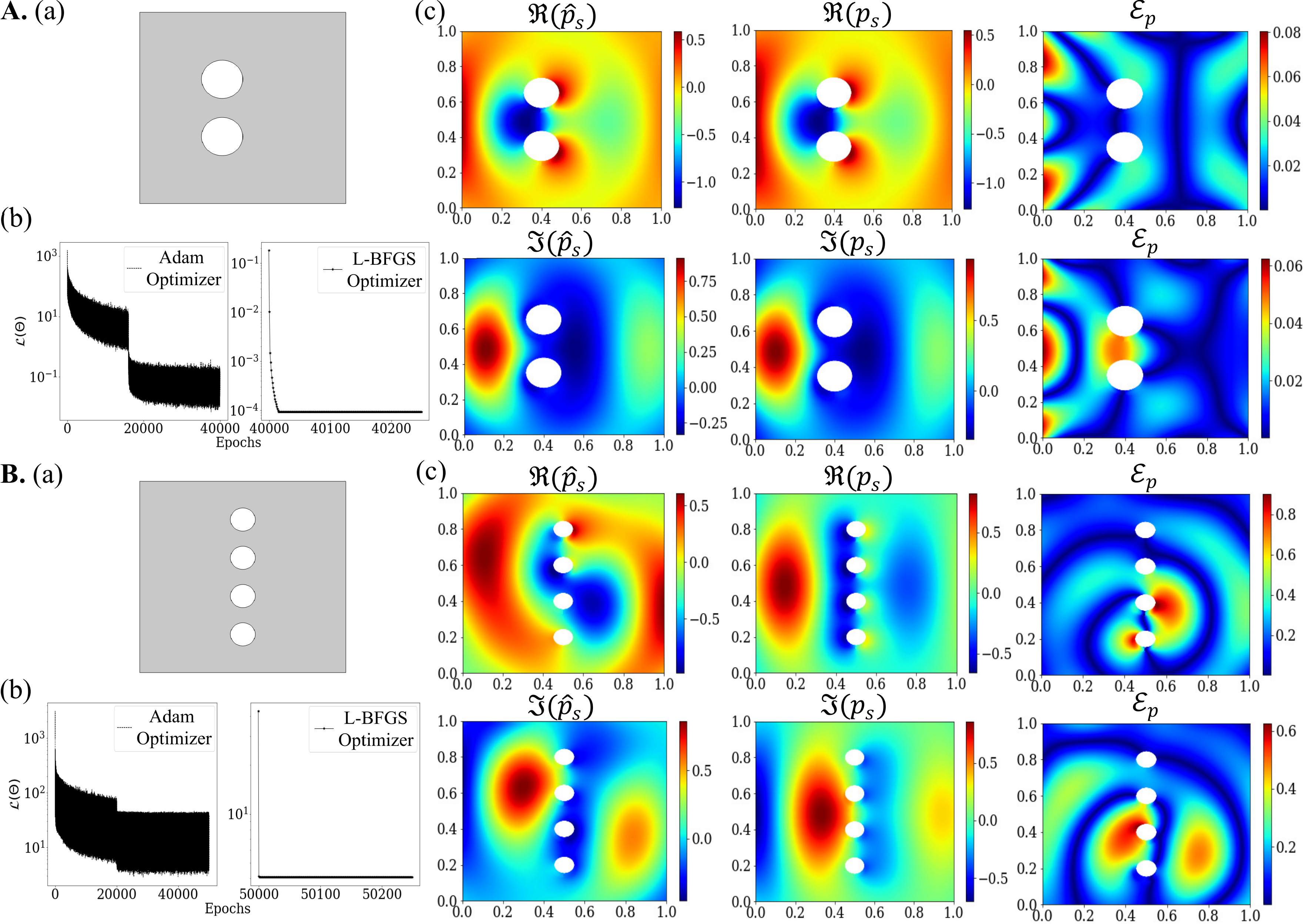}
	\caption{\textbf{Case 3- $b$-PINN:} Illustration of the performance analysis of $b$-PINN for multiple scattering applications in (A)-(B). The simulations in (A) and (B) are performed for planar wavefields at $f=500~Hz$ incident on four scatterers and four scatterers, respectively. In each case, schematic: (a) Represents the geometry of the acoustic domain embedded with a rigid scatterer. (b) Plots the variation in the loss function with training epochs. Note that the plots include the loss variation using both Adam and L-BFGS optimizers. (c) Amplitude maps of the real (top) and imaginary (bottom) components of the predicted pressure field ($\hat{p}_s$), true pressure field ($p_s$), and the point-wise error ($\mathcal{E}$) between $\hat{p}_s$ and $p_s$.}
	\label{fig: MultiSc_BP}
\end{figure}
The schematics in Fig.~\ref{fig: MultiSc_BP} illustrate the prediction accuracy of the $b$-PINN for multiple scattering simulations. Fig.~\ref{fig: MultiSc_BP}(A) highlights the prediction assessment for a multiple scattering problem with two scatterers. Here, in addition to the network parameters represented in Table~\ref{table: Table_2}, the $b$-PINN is trained for $N_e=40,000$ epochs with a step learning rate of $l_r=1e-3$ for the first $20,000$ epochs and $l_r=1e-4$ for the rest of the training with Adam Optimizer (Fig.~\ref{fig: MultiSc_BP}(A(b))). The comparison between $\hat{p}_s$ and $p_s$ in Fig.~\ref{fig: MultiSc_BP}(A) evaluates $[L_2=0.072, R^2=0.9942]$, indicating a good prediction accuracy of the $b$-PINN for a multiple scattering problem with two scatterers. Further, the $b$-PINN is trained to simulate a multiple scattering problem with four scatterers embedded in $\Omega_0$ as shown in Fig.~\ref{fig: MultiSc_BP}(B(a)). Here, the $b$-PINN is trained for $N_e=50,000$ epochs with a step learning rate of $l_r=1e-3$ for the first $20,000$ epochs and $l_r=1e-4$ for the rest of the training with Adam Optimizer. A direct comparison between the $\hat{p}_s$ and $p_s$ fields in Fig.~\ref{fig: MultiSc_BP}(B(c)) presents low accuracy in the $b$-PINN approximation for multiple scattering with four scatterers. Moreover, the error metrics values $[L_2=0.2816, R^2=0.8319]$, also indicate a high prediction error. This inaccuracy can be directly attributed to the inability of the $b$-PINN to handle the specific four scatterers configuration. More specifically, the $b$-PINN finds it challenging to learn to approximate the simulated field due to four scatterers as indicated by the fact that the value of the training loss reaches a plateau (see Fig.~\ref{fig: MultiSc_BP}(B(b))). A subsequent analysis reveals that this high prediction error is directly caused by the rigid scatterers boundaries as indicated by the high $\mathcal{L}_{\mathcal{B}_i}(\theta)$ values. 
This observation reinforces our assumption that as the number of scatterers increases, the high prediction error observed during the initial network training phase propagates across the spatial domain throughout the entire training process. Therefore, while the $b$-PINN demonstrates good performance and prediction accuracy when simulating a few scatterers, it encounters limitations when solving multiple scattering problems with an increasing number of scatterers.

\subsubsection{Multiple scattering simulation using $s$-PINN}

\begin{table}[ht]
    \centering
    \begin{tabular}{|c|c|c|c|c|c|c|c|c|}
        \hline
        \multicolumn{1}{|c|}{\textbf{Parameter}}  & \multicolumn{4}{c|}{\textbf{Case 3 $s$-PINN}} & \multicolumn{4}{c|}{\textbf{Case 3 $s$-PINN}} \\
        \multicolumn{1}{|c|}{$4 \times 4$ lattice}  & \multicolumn{4}{c|}{($f=500~Hz$)} & \multicolumn{4}{c|}{($f=1~kHz$)} \\  \cline{2-9}
         &   PINN$^1$ & PINN$^2$ & PINN$^3$ & PINN$^4$ & PINN$^1$ & PINN$^2$ & PINN$^3$ & PINN$^4$ \\
        \hline
        \hline
		$n_w$ & 100 & 100 & 100 & 100 & 125 & 125 & 125 & 125\\
        \hline   
		$n_r$ & 2 & 2 & 2 & 2 & 3 & 3 & 3 & 3\\
		\hline        
		$n_l$ & 4 & 4 & 4 & 4 & 4 & 4 & 4 & 4\\
		\hline
		$N_{b_i}$ & 100 & 100 & 100 & 100 & 100 & 100 & 100 & 100\\
        \hline
   	$N_{b_e}$ & 1000 & 1000 & 1000 & 
        1000 & 1000 & 1000 & 1000 & 1000\\
        \hline
        $N_n$ & 10,000 & 10,000 & 10,000 & 10,000 & 10,000 & 10,000 & 10,000 & 10,000 \\
        \hline
        \multicolumn{1}{|c|}{$l_r$}  & \multicolumn{4}{c|}{1e-3, 1e-4} & \multicolumn{4}{c|}{1e-3, 1e-4} \\ 
        \hline
    \end{tabular}
    \caption{Summary of key network parameters used to develop the $s$-PINN architecture for multiple scattering simulation due to a $4 \times 4$ scatterer lattice at $f=500~Hz$ and $f=1~kHz$.}
    \label{table: Table_3}
\end{table}

\begin{figure}
	\centering
	\includegraphics[width=1.0\linewidth]{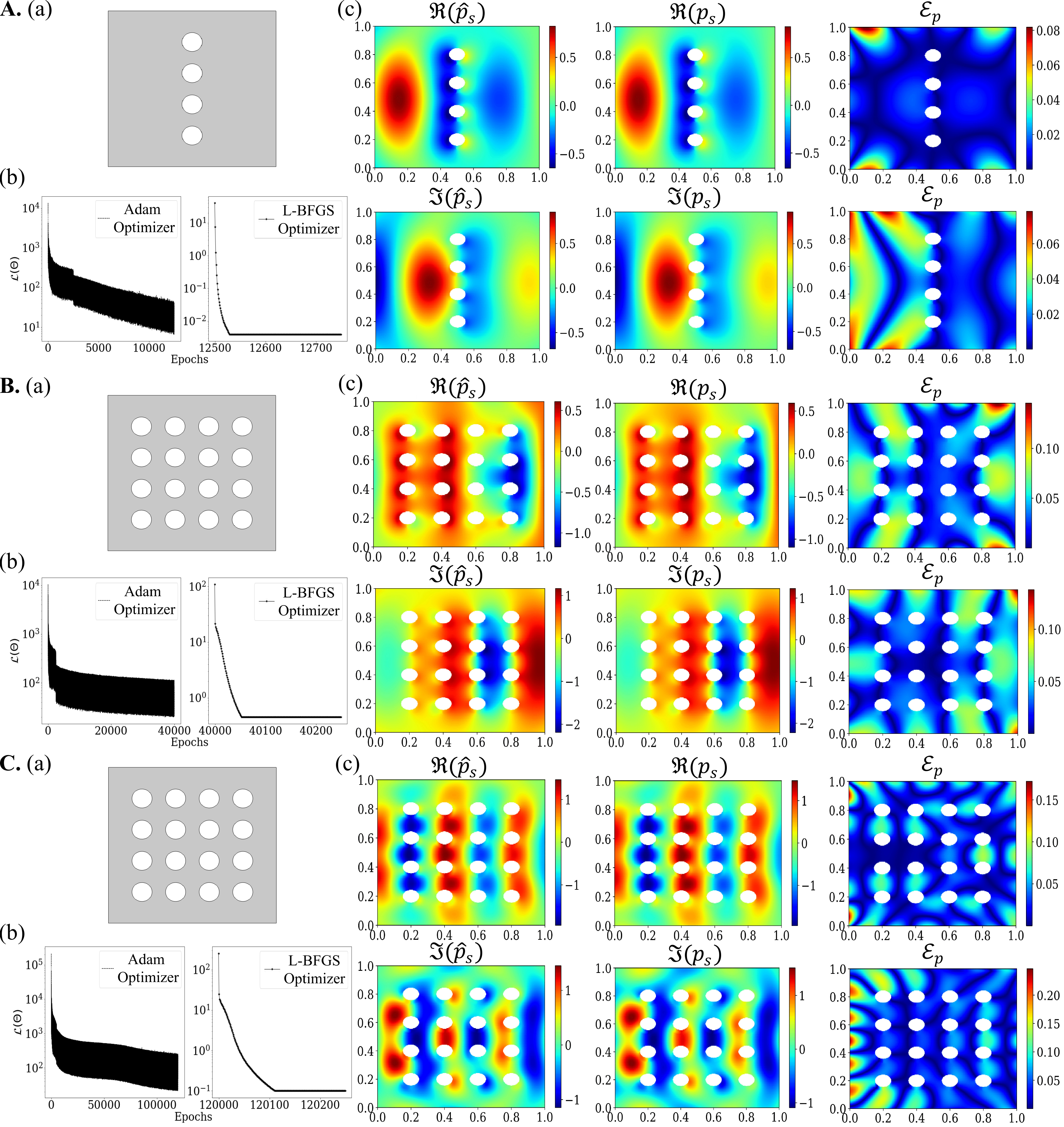}
	\caption{\textbf{Case 3- $s$-PINN:} Illustration of the performance analysis of $s$-PINN for multiple scattering applications in (A)-(C). The simulations in (A), (B), and (C) are performed for planar wavefields incident on scatterer lattices with scatterers arranged in $4 \times 1$ grid at $f=500~Hz$, $4 \times 4$ grid at $f=500~Hz$, and $4 \times 4$ grid at $f=1~kHz$, respectively. In each case, schematic: (a) represents the geometry of the acoustic domain embedded with a rigid scatterer. (b) Plots the variation in the loss function with training epochs. Note that the plots include the loss variation using both Adam and L-BFGS optimizers. (c) Amplitude maps of the real (top) and imaginary (bottom) components of the predicted pressure field ($\hat{p}_s$), true pressure field ($p_s$), and the point-wise error ($\mathcal{E}$) between $\hat{p}_s$ and $p_s$.}
	\label{fig: MultiSc_SP}
\end{figure}
As the $s$-PINN is designed to address the limitations of the $b$-PINN in the context of multiple scattering simulations, we first compare the performance of the two networks (i.e. $s$-PINN and $b$-PINN). Specifically, we compare the performance of the $b$-PINN (see Fig.~\ref{fig: MultiSc_BP}(B)) and $s$-PINN (see Fig.~\ref{fig: MultiSc_SP}(A)) for a $4 \times 1$ acoustic lattice simulated at $f=500~Hz$. Similar to the other schematics shown in this section, the (a), (b), and (c) parts of Fig.~\ref{fig: MultiSc_BP}(B) for $b$-PINN and Fig.~\ref{fig: MultiSc_SP}(A) for $s$-PINN represent the simulated domain, loss variation, and network prediction accuracy, respectively.
A direct comparison of the variation in training losses between the $b$-PINN and the $s$-PINN highlights the ability of the $s$-PINN to learn to approximate the scattered fields for multiple scattering simulations. Further, the $s$-PINN predictions return $[L_2=0.0694, R^2=0.9946]$, which represents a $75\%$ increase in the prediction accuracy compared to the $b$-PINN prediction for the same simulation domain. 

Note that Table~\ref{table: Table_2} presents the values of key network parameters selected for each $b$-PINN within the $s$-PINN to simulate the $4 \times 1$ scatterer lattice. Here, the $s$-PINN is trained with a total number of $N'_s=4$ scatterers using $N_s=4$ $b$-PINNs each with $n_s=1$ scatterer. Moreover, this $s$-PINN is trained for $N_e=12,500$ epochs with a step learning rate of $l_r=1e-3$ for the first $2500$ epochs and $l_r=1e-4$ for the rest of the training with Adam Optimizer. Although the individual $b$-PINN within the $s$-PINN is designed with lower $n_l$ and $N_e$ compared to when trained with a single $b$-PINN, the much higher prediction accuracy of the $s$-PINN indicates the remarkable ability of the network architecture to enforce the superposition principle along with the minimization of the governing PDE and boundary conditions on individual scatterer sub-domains.

As the above results established the superior performance of the $s$-PINN over the $b$-PINN for multiple-scattering simulations, in the following we investigate the ability of $s$-PINN to simulate more complex multiple scattering cases. Table~\ref{table: Table_3} records the values of the key network parameters used to develop the $s$-PINN for a more elaborate scattering scenario. Specifically, Fig.~\ref{fig: MultiSc_SP}(B) indicates the network performance assessment for a $4 \times 4$ lattice with a total number of $N'_s=16$ scatterers simulated at $f=500~Hz$ by using $N_s=4$ $b$-PINNs each with $n_s=4$ scatterers. The comparison of the predictions and ground truth in Fig.~\ref{fig: MultiSc_SP}(B(c)), and of the metrics $[L_2=0.1076, R^2=0.9861]$ highlights the good prediction accuracy of the $s$-PINN. Moreover, this $s$-PINN is trained for $N_e=40,000$ epochs with a step learning rate. In addition, we also simulate the $4 \times 4$ scatterer lattice at $f=1~kHz$ and study the performance at higher frequencies as shown in Fig.~\ref{fig: MultiSc_SP}(C) with a step learning rate for $N_e=120,000$ epochs. Based on the prediction comparison in Fig.~\ref{fig: MultiSc_SP}(C(c)), the accuracy metrics are evaluated as $[L_2=0.0883, R^2=0.9913]$. The low $L_2$-error and high $R^2$-score highlight the performance of $s$-PINN when simulating complex multiple scattering problems at different frequencies in the low- and mid-frequency ranges. Note that as we move into the high-frequency range, the prediction accuracy of the current $s$-PINN begins to deteriorate due to the inability of the current network to capture finer wavefield features with existing spatial sampling. Therefore, although the proposed network model has lower discretization dependence in comparison to mesh-based approaches like FEM, the high-frequency simulations require finer discretization to capture more accurate results.

\section{Conclusions}
\label{ssec: Conclusion}

This study presented a physics-informed machine learning framework designed to simulate 2D acoustic scattering from arbitrary-shaped objects embedded in an unbounded domain. The framework encompassed two distinct physics-informed neural network (PINN) models, referred to as the baseline-PINN ($b$-PINN) and the superposition-PINN ($s$-PINN). The $b$-PINN was developed to perform acoustic scattering simulations when in presence of either a single or a small number of rigid scatterers. The two key attributes of the $b$-PINN are its ability to simulate the scattered field due to arbitrary-shaped rigid scatterers and to address acoustic scattering problems with incident wavelengths shorter than the characteristic scatterer sizes (i.e. towards the geometric scattering regime). However, while the $b$-PINN can handle multiple scattering, it was shown that its applicability is restricted to scenarios with only a few rigid scatterers due to error propagation. To overcome this limitation, an alternative configuration, denominated the superposition PINN, was developed. 

While the $s$-PINN builds upon the basic $b$-PINN architecture, it overcomes the error propagation issue by using a domain partitioning approach. Multiple copies of the $b$-PINN used to simulate individual parts of the domain and then they are combined together by means of an architecture that embodies the superposition principle of linear acoustics. In addition to its ability to simulate multiple scattering problems, the results highlighted the discretization-independent nature of the proposed network that enables handling simulations of arbitrarily shaped scatterers and a wide range of frequencies using the same spatial discretization. Furthermore, the capability of $s$-PINN to facilitate parallel computations by assigning sub-domain simulations to separate computational nodes can significantly reduce the total computational time for complex multiple scattering simulations. These properties propel the $s$-PINN as a capable forward solver that can address the challenges associated with the high computational cost in traditional mesh-based methods like finite element method (FEM) for complex multiple scattering simulations.  

In contrast to the existing data-driven DNNs, the proposed PINN framework is trained by explicitly enforcing the governing equations and boundary conditions of the problem. This approach enables it to generate accurate and physically consistent solutions without relying on labeled datasets for network training, thereby reducing the data generation cost by 100\%. Furthermore, the performance of the proposed network models was evaluated by comparing their predictions to FEM ground truth for various acoustic scattering applications. The numerical investigations highlighted the ability of the network models to accurately simulate acoustic scattering scenarios across applications involving arbitrary scatterer shapes, high frequencies, and lattices of scatterers.
Additional areas needing more research and development exist in the $s$-PINN architecture. As an example, $s$-PINN has the potential to be set up and trained using parallel computations, therefore its implementation and performance should be explored. In addition, the current $s$-PINN architecture becomes ineffective at capturing multi-scatterered fields with nonlinear behavior due to substantially higher acoustic pressure amplitudes, e.g., shock waves. Therefore, developing a PINN architecture to handle multiple scattering phenomenon while accommodating the corresponding nonlinear behavior is an interesting area of potential future research.

\bigskip

\noindent \textbf{Competing interests}

The authors declare no competing interest.\\

\noindent \textbf{Acknowledgments}

This work was supported by the Laboratory Directed Research and Development program at Sandia National Laboratories, a multimission laboratory managed and operated by National Technology and Engineering Solutions of Sandia LLC, a wholly owned subsidiary of Honeywell International Inc. for the U.S. Department of Energy’s National Nuclear Security Administration under contract DE-NA0003525.

\bibliographystyle{unsrt} 
\bibliography{Finaldraft}

\end{document}